\documentclass[conference]{IEEEtran}
\IEEEoverridecommandlockouts
\usepackage{cite}
\usepackage{amsmath,amssymb,amsfonts}
\usepackage{algorithmic}
\usepackage{graphicx}
\usepackage{textcomp}
\usepackage{xcolor}
\usepackage{booktabs}
\def\BibTeX{{\rm B\kern-.05em{\sc i\kern-.025em b}\kern-.08em
    T\kern-.1667em\lower.7ex\hbox{E}\kern-.125emX}}

\begin{document}

\title{Human Blastocyst Classification after In Vitro Fertilization Using Deep Learning}

\IEEEpubid{978-1-7281-8038-0/20/\$31.00~\copyright~2020 IEEE}
\IEEEpubidadjcol

\author{\IEEEauthorblockN{Ali Akbar Septiandri$^1$, Ade Jamal$^2$}
\IEEEauthorblockA{Faculty of Science and Technology \\
Universitas Al Azhar Indonesia \\
Jakarta, Indonesia \\
$^1$aliakbars@live.com, $^2$adja@uai.ac.id}
\and
\IEEEauthorblockN{Pritta Ameilia Iffanolida, Oki Riayati, Budi Wiweko}
\IEEEauthorblockA{Human Reproductive, Infertility, and Family Planning Research Center IMERI \\
Faculty of Medicine \\
Universitas Indonesia \\
Jakarta, Indonesia}
}

\maketitle

\begin{abstract}
Embryo quality assessment after in vitro fertilization (IVF) is primarily done visually by embryologists. Variability among assessors, however, remains one of the main causes of the low success rate of IVF. This study aims to develop an automated embryo assessment based on a deep learning model. This study includes a total of 1084 images from 1226 embryos. The images were captured by an inverted microscope at day 3 after fertilization. The images were labelled based on Veeck criteria that differentiate embryos to grade 1 to 5 based on the size of the blastomere and the grade of fragmentation. Our deep learning grading results were compared to the grading results from trained embryologists to evaluate the model performance. Our best model from fine-tuning a pre-trained ResNet50 on the dataset results in 91.79\% accuracy. The model presented could be developed into an automated embryo assessment method in point-of-care settings.
\end{abstract}


\section{Introduction}

Embryo quality plays a pivotal role in a successful IVF cycle. Embryologists assess embryo quality from the morphological appearance using direct visualization \cite{cummins1986formula}. There are three protocols in different time points to evaluate the quality of an embryo: (1) quality assessment of zygote (16-18 hours after oocyte insemination), (2) morphological quality assessment of cleavage stage embryos (day 3 after insemination), and (3) Morphological quality assessment of blastocyst stage embryos (4-5 days after fertilization) \cite{nasiri2015overview}.

This kind of visual assessment is susceptible to the subjectivity of the embryologists. There are two kinds of variability in embryo assessment as seen in \cite{bendus2006interobserver}: interobserver and intraobserver. Grading systems like Veeck criteria \cite{veeck1999atlas} aim to standardize grading and minimize both variabilities. However, as also found in \cite{bendus2006interobserver}, ``the embryologists often gave an embryo a score different than Dr. Veeck, but that score was typically within one grade.'' The study also shows that the intraobserver variation is limited. Khosravi et al. \cite{khosravi2019deep} also shows that only 89 out of 394 embryos were classified as the same quality by all five embryologists in their study.

In recent years, we have applications of deep learning for computer vision in medical imaging to address this variability issue. From MRI for brain imaging \cite{akkus2017deep}, various anatomical areas \cite{litjens2017survey}, to point-of-care diagnostics from microscopy images \cite{quinn2016deep}, deep learning has aided medical practitioners to diagnose better. In the field of reproductive medicine, we have also seen some application of artificial intelligence as shown in \cite{zaninovic2019artificial}. Recent studies also explored the possibilities to automate embryo assessments for IVF \cite{khosravi2019deep,kragh2019automatic,chen2019using} using a robust classifier trained on thousands of images.

Our main contribution is that previous studies \cite{khosravi2019deep,kragh2019automatic,chen2019using} are using day 5 embryo images, while we are using day 3 embryo images. As seen in \cite{coskun2000day}, ``Early embryos can grow in a simple salt solution, whereas they require more complex media after they reach the 8-cell stage.'' Day 3 and day 5 embryos are similar in implantation, clinical pregnancy, twinning, and live birth rates \cite{hatirnaz2017day,coskun2000day}, but since day 5 embryos are extremely sensitive to suboptimal culture environment, many clinics are still doing day 3 embryo transfers \cite{lan2003predictive}. Moreover, unlike previous studies which only used ResNet50, we also compared several deep learning architectures to do the task. The dataset used in this study is described in the following section.

\IEEEpubidadjcol

\section{Dataset}

Our dataset comprises of 1084 images from 1226 embryos of 246 IVF cycles at Yasmin IVF Clinic, Jakarta, Indonesia. The images were captured by an inverted microscope at day 3 after fertilization. The images consist of 2-3 embryos each. We manually cropped the images and a team of 4 embryologists graded them 1-5 by using Veeck criteria \cite{veeck1999atlas}. However, we only found grade 1 to grade 3 embryos from the samples. This yields 1226 identified embryos consisting of 459 grade 1, 620 grade 2, and 147 grade 3 embryos.

\begin{figure}[htbp]
    \centering
    \includegraphics[width=.8\columnwidth]{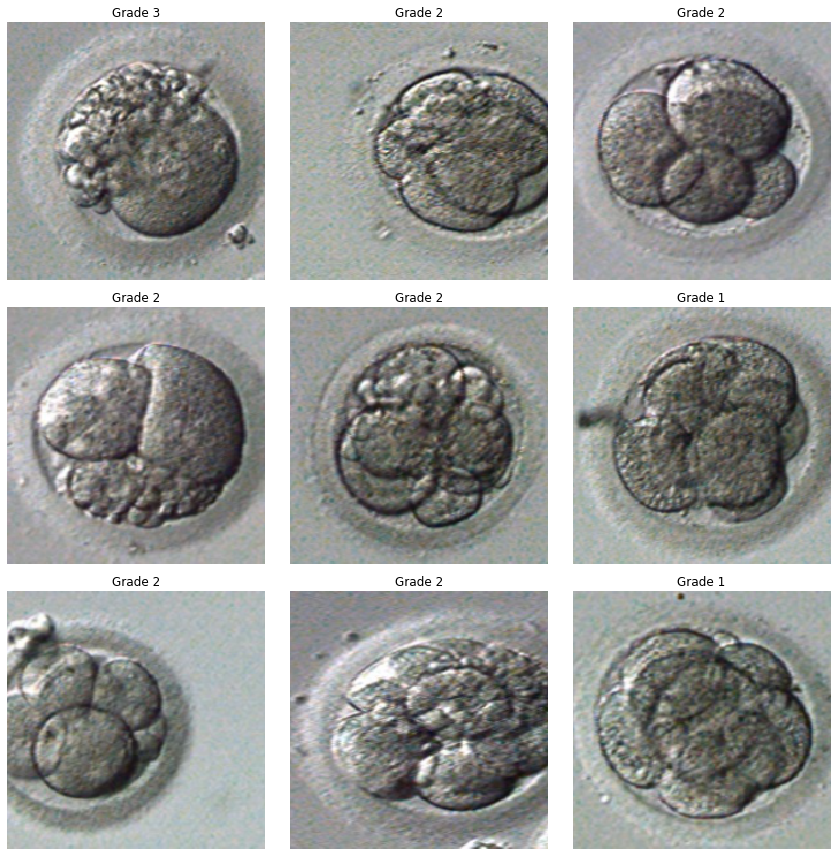}
    \caption{Embryo images after the preprocessing steps}
    \label{fig:examples}
\end{figure}

To train the model and for further generalization error, we divided the dataset into train and test sets with 75:25 ratio. The training set is further divided into training and validation sets with 70:30 ratio. Some examples of the images in the training set can be seen in Figure~\ref{fig:examples}. These images were automatically cropped and resized by the library that we used for the deep learning application \cite{howard2020fastai}. This preprocessing step was done to prevent overfitting of the models.

\section{Methodology}

We used the fast.ai library \cite{howard2020fastai} for the deep learning application. The library helped us to do transfer learning from several pre-trained convolutional neural networks \cite{oquab2014learning}, such as the residual networks \cite{he2016deep} with different depths (ResNet18, ResNet34, ResNet50, ResNet101), densely connected convolutional networks \cite{huang2017densely}, Xception \cite{chollet2017xception}, and the MobileNetV2 \cite{sandler2018mobilenetv2} to the given task. We trained these models using backpropagation with 1cycle policy \cite{smith2017cyclical}. We used the cyclical learning rates \cite{smith2017cyclical} implemented in the fast.ai library to find the best learning rates. To find an unbiased estimate of the accuracy from a model, we trained the model and predicted the test set five times after ensuring that we got the best model during the training step.

\subsection{Deep Residual Networks}

Prior to the study on residual learning, deeper neural networks are harder to train \cite{he2016deep}. The accuracy of deeper neural networks gets saturated and becomes worse eventually. Their solution is to recast the original mapping $\mathcal{H}(\mathbf{x})$ into $\mathcal{F}(\mathbf{x}) + \mathbf{x}$ that can be seen as a feedforward neural network with shortcut connections. This reformulation makes it easier to optimize the model and can even reach over 1000 layers with no optimization difficulty though achieved worse result compared to the ones with fewer layers. The ensemble of this architecture has 3.57\% top-5 error on the ImageNet test set and also won the 1st places in several tracks in ILSVRC and COCO 2015 competitions.

\subsection{Densely Connected Convolutional Networks}

Borrowing ideas from deep residual networks, the Dense Convolutional Network (DenseNet) tries to harness the power of shortcut connections. ``For each layer, the feature-maps of all preceding layers are used as inputs, and its own feature-maps are used as inputs into all subsequent layers.'' \cite{huang2017densely} This enables the model to have ``substantially fewer parameters and less computation'' while still achieving state-of-the-art performances. Nevertheless, DenseNet can still be scaled into hundreds of layers easily.

\subsection{Xception}

We also used the Xception architecture \cite{chollet2017xception} to benchmark against the study in \cite{kragh2019automatic}. This architecture is an improvement from the Inception V3 \cite{szegedy2016rethinking} ``where Inception modules have been replaced with depthwise separable convolutions''. Xception also uses residual connections as in the deep residual networks \cite{he2016deep}. It outperforms Inception V3 while having a similar number of parameters as Inception V3.

\subsection{MobileNetV2}

Since the result would possibly be implemented in a mobile phone for better outreach, we consider MobileNetV2 as a viable architecture. This architecture allows us to ``reduce the memory footprint needed during inference by never fully materializing large intermediate tensors.'' \cite{sandler2018mobilenetv2}

\section{Results and Discussion}

We found that the best learning rates are around $5 \times 10^{-3}$ from 8 epochs. After 8 epochs, the learning curve starts plateauing which suggest the model overfits the training data. The results from all models can be seen in Table~\ref{tab:results}. We can see that the ResNet50 got the highest accuracy and the lowest cross-entropy loss of all models. While increasing the depth of the ResNet model from 18, 34, to 50 increased the accuracy, it stopped increasing afterwards. We argue that this is because the dataset is relatively simpler compared to ImageNet where we have different objects in different colours and sizes. This might also be the case why DenseNet models failed to achieve better performance while being more complex.

\begin{table}[htbp]
    \centering
    \caption{Model comparison}
    \begin{tabular}{ccc}
    \toprule
    model           &             accuracy &       loss \\
    \midrule
    ResNet18        & $89.38\% \pm 0.75\%$ & $0.3312 \pm 0.0330$ \\
    ResNet34        & $89.97\% \pm 1.27\%$ & $0.3495 \pm 0.0343$ \\
    ResNet50        & $\mathbf{91.79\% \pm 0.48\%}$ & $\mathbf{0.3114 \pm 0.0253}$ \\
    ResNet101       & $91.07\% \pm 1.00\%$ & $0.3749 \pm 0.0623$ \\
    DenseNet121     & $89.97\% \pm 0.27\%$ & $0.3567 \pm 0.0365$ \\
    DenseNet169     & $91.14\% \pm 0.54\%$ & $0.3472 \pm 0.0366$ \\
    Xception        & $88.86\% \pm 0.96\%$ & $0.3209 \pm 0.0206$ \\
    MobileNetV2     & $91.14\% \pm 0.84\%$ & $0.3442 \pm 0.0258$ \\
    \bottomrule
    \end{tabular}
    \label{tab:results}
\end{table}

\begin{figure}[htbp]
    \centering
    \includegraphics[width=\columnwidth]{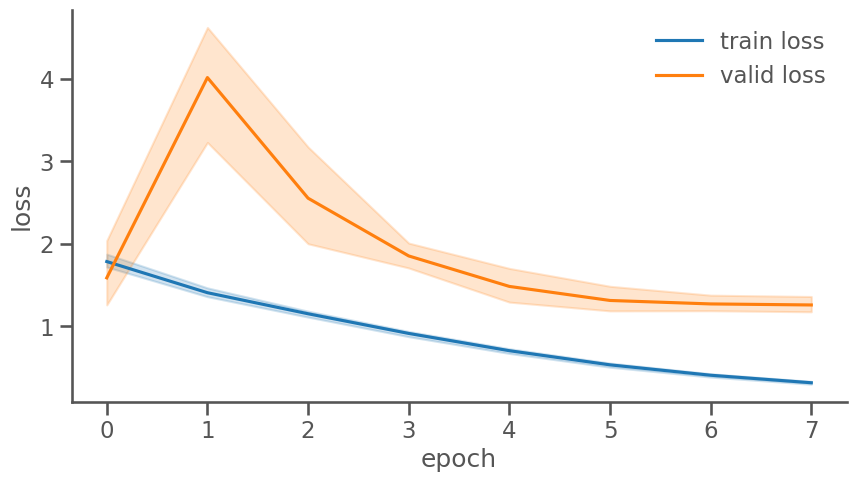}
    \caption{Learning curve from MobileNetV2}
    \label{fig:learning_curve}
\end{figure}

\begin{figure}[htbp]
    \centering
    \includegraphics[width=.6\columnwidth]{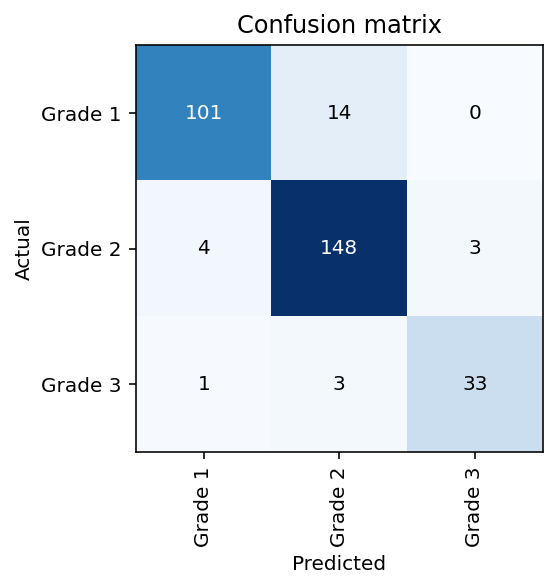}
    \caption{A confusion matrix from MobileNetV2}
    \label{fig:conf_matrix}
\end{figure}

On the other hand, we are more interested in MobileNetV2 which achieved a similar accuracy to the best model. As we elaborated in the previous section, this architecture would enable us to design an embedded system for point-of-care diagnostics. Thus, we provide the learning curve from MobileNetV2 in Figure~\ref{fig:learning_curve}. An example of a confusion matrix from MobileNetV2's prediction can be seen in Figure~\ref{fig:conf_matrix}.

Note that we are only using three embryo grades in this study due to the unavailability of the samples. We would need to reassess these models when we have more grades. However, since the current model can predict the minority class (grade 3) well, we argue that it might generalise with the complete grades as well.

\begin{figure}[htbp]
    \centering
    \includegraphics[width=.8\columnwidth]{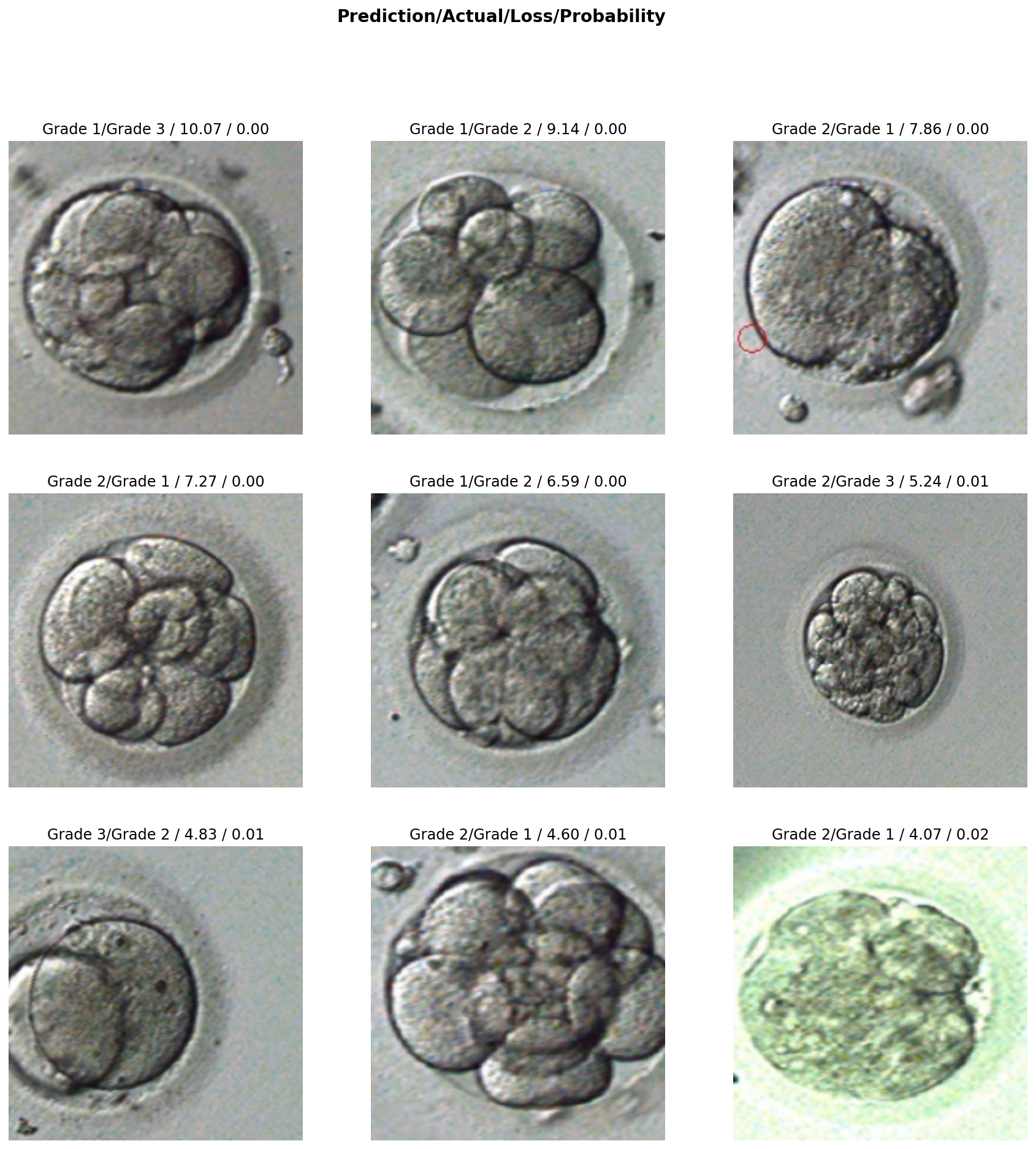}
    \caption{Examples of misclassified images}
    \label{fig:misclassified}
\end{figure}

Examples from the misclassified embryos as seen in Figure~\ref{fig:misclassified} suggest that the image capturing process can impact the model performance. For example, different shades of color (bottom right image) of the embryo images might cause the misclassification. Obstruction in the images, such as the red circles (top right image) or timestamps from the application that we used to digitally process the microscopy images also affect the performance of the models.

\section{Related Work}

Robust assessment and selection of human blastocysts after IVF using deep learning has been studied in \cite{khosravi2019deep,kragh2019automatic,chen2019using}. Khosravi et al. \cite{khosravi2019deep} fine-tuned their Inception V1 model \cite{szegedy2015going} on 10,148 images of day 5 embryos from the Center for Reproductive Medicine at Weill Cornell Medicine. The accuracy of the resulting model is 96.94\%. However, the training was done for good and poor quality images only from the three classes defined first.

Kragh et al. \cite{kragh2019automatic} address the issue in \cite{khosravi2019deep} by predicting blastocyst grades of any embryo based on raw time-lapse image sequences. Aside from using Xception \cite{chollet2017xception} as the convolutional neural network (CNN) architecture to extract image features, they also use a recurrent neural network (RNN) ``that connects subsequent frames from the sequence in order to leverage temporal information.'' They predict the inner cell mass (ICM) and trophectoderm (TE) grades for the entire sequence from the RNN using two independent fully-connected layers. They train the models on 6957 embryos. On a test set of 55 embryos annotated by multiple embryologists, their models reached 71.9\% and 76.4\% of ICM and TE accuracy respectively compared to human embryologists who only achieved 65.1\% and 73.8\%. While the result is promising, using a RNN makes the training slower and prone to vanishing or exploding gradients \cite{pascanu2013difficulty}.

In \cite{chen2019using}, the authors fine-tune a ResNet50 model on 171,239 images from 16,201 day 5 embryos to predict blastocyst development ranking from 3–6, ICM quality, and TE quality. The images were annotated by embryologists based on Gardner's grading system. They achieved ``an  average  predictive  accuracy  of  75.36\%  for  the  all  three  grading categories: 96.24\% for blastocyst development, 91.07\% for ICM quality, and 84.42\% for TE quality.''

\section{Conclusions}

We have shown in this study that we can grade day 3 embryo images automatically with the best accuracy of 91.79\% by fine-tuning a ResNet50 model. We found that more complex models failed to achieve better accuracy compared to the ResNet50. MobileNetV2 as our model of interest to build an embedded system achieved a relatively similar accuracy of 91.14\% compared to the best model. The models still face some problems from different shades of colour or obstructions from the software that embryologists use to capture and process the images.

We saw good results when combining CNN and RNN in previous studies. However, in resource constrained settings, e.g. when we want to make inferences on small devices, the unparallelizable nature of RNNs also makes it challenging to implement. Moreover, time-lapse microscopes are not prevalent in developing countries. Thus, our solution would be more feasible to put into production.

Since we are still manually cropping the embryos from the original images, we can extend this work to automate this task, e.g. using an image segmentation model like YOLOv3 \cite{redmon2018yolov3} or U-Net \cite{ronneberger2015u}. In the future, we hope that this model can be developed as an embedded system for point-of-care diagnostics such as found in \cite{quinn2016deep}.




\bibliographystyle{IEEEtran}
\bibliography{embryo}

\end{document}